# The Thiocyanate Anion is a Primary Driver of Carbon Dioxide Capture by Ionic Liquids

Vitaly Chaban[1]

MEMPHYS — Center for Biomembrane Physics, Syddansk Universitet, Odense M., 5230, Kingdom of Denmark

**Abstract**. Carbon dioxide, $CO_2$, capture by room-temperature ionic liquids (RTILs) is a vivid research area featuring both accomplishments and frustrations. This work employs the PM7-MD method to simulate adsorption of $CO_2$ by 1,3-dimethylimidazolium thiocyanate at 300 K. The obtained result evidences that the thiocyanate anion plays a key role in gas capture, whereas the impact of the 1,3-dimethylimidazolium cation is mediocre. Decomposition of the computed wave function on the individual molecular orbitals confirms that $CO_2$-SCN binding extends beyond just expected electrostatic interactions in the ion-molecular system and involves partial sharing of valence orbitals.

**Key words**: ionic liquid, carbon dioxide, thiocyanate, PM7-MD.

[1] E-mail: vvchaban@gmail.com

**TOC Image**

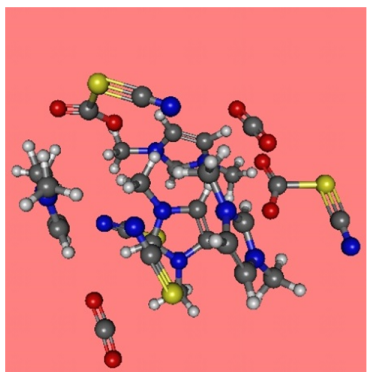

**Introduction**

Room-temperature ionic liquids (RTILs) play an important role in a variety of chemical processes and setups.[1-13] The tunability of RTILs allows for achieving goals, which were previously inaccessible for conventional molecular liquids. New RTILs are reported continuously fostering new ideas and applications. Not all available ionic liquids are RTILs though.[14] Some of them exhibit quite high melting temperatures due to relatively favorable packing in the cation-anion lattice and strong electrostatic attraction.[14]

Carbon dioxide, $CO_2$, is routinely produced in great quantities upon combustion of fossil fuels. Excess of $CO_2$ in the atmosphere was recently realized to be a serious threat for sustainable development of our civilization. The major problem is global warming, which at least partially correlates with the excessive emission of $CO_2$. The technologies for $CO_2$ removal have attracted considerable attention in the research community. In the carbon capture and storage technology, widely known as CCS, carbon dioxide is captured via physical or chemical adsorption or, sometimes, via membrane separation.[15] Aqueous amine compounds represent the major working horse in the $CO_2$ capture stage. This stage also includes separation of carbon dioxide from other omnipresent gases, such as nitrogen, oxygen, water vapor, noble gases, etc.[16-18]

Despite being definitely useful and widely applied, CCS is not cost efficient. Regeneration of adsorbent after $CO_2$ removal requires large amounts of energy. Furthermore, certain amine adsorbents are pyrolyzed upon heating and need continuous replacement. Active research is being conducted presently to develop more energetically efficient $CO_2$ capturing systems or to locate new working solutions.[19, 20]

Various ionic liquids[13, 20-22] were probed for carbon dioxide capture, although with changeable success. Unlike conventional ionic compounds, RTILs possess both polar and nonpolar moieties within a single ion pair. They can capture gases not only due to stronger dipole-dipole binding, but also due to a weaker van der Waals interaction. Being liquid at working temperatures, RTILs can capture $CO_2$ thanks to a favorable entropic contribution to free energy of mixing. This is in addition to favorable electrostatic attraction (enthalpy gain). An ability to combine various cations and anions within a single substance provides a great opportunity to tune desirable gas capture behavior of a particular RTIL. It was experimentally shown recently[19, 23] that RTILs do exhibit selectivity towards $CO_2$ in the presence of $O_2$, $N_2$ and gaseous organic admixtures, such as methane and acetylene. RTILs swallow $CO_2$ preferentially via physical adsorption, which implies smaller energy expenses at the solvent regeneration stage. Adsorption of molecular compounds unlikely can significantly alter volatility of RTILs. Therefore, solvent loss is essentially avoided.

Ionic liquids, which contain thiocyanide, $SCN^-$ anions, represent a high interest due to our expectations of specific non-bonded interactions between $CO_2$ and $SCN^-$. These expectations are partially based on the existing experimental evidence of a good $CO_2$ capturing ability, obtained using a high-pressure variable-volume view cell.[19] Our recently introduced PM7-MD methodology is applied to simulate $CO_2$ adsorption and structure in 1,3-dimethylimidazolium thiocyanate, [MMIM][SCN]. PM7-MD provides electronic structure level physical insights for heated molecular systems (entropy effect included) at a reasonably inexpensive computational cost.

**Methodology**

The provided analysis is based on the molecular dynamics trajectories, which were recorded using an electronic structure description of the nuclear-electronic system. The wave function of all given molecular configurations was optimized repeatedly using the self-consistent field methods (SCF, convergence criterion of $10^{-7}$ Ha). The immediate force acting on each atomic nucleus was derived from the optimized wave function at each time-step. The nuclear trajectory was propagated fully classically via the velocity Verlet integration scheme. Both velocity and position were calculated at the same value of the time variable. The applied simulation scheme obeys the Born-Oppenheimer approximation, which is widely applied in numerical simulations within the fields of theoretical chemistry and molecular physics.

The wave function at each nuclear time-step was optimized using the PM7 ("parametrized model seven") semiempirical Hamiltonian.[6, 24, 25] PM7 is a quantum chemistry method, which is based on the Hartree-Fock (HF) formalism. It is, therefore, more theoretically fundamental and robust for applications, as compared to empirical force field methods. PM7 was responsible for the electronic part of the calculation to obtain molecular orbitals, heat of formation, and its derivative with respect to molecular geometry. However, as compared to the non-parametrized HF method, PM7 makes a set of approximations and adopts certain parameters in its equations from experimental data to speed up calculations and facilitate wave function convergence.[6, 24, 25] Except serving for faster performance of SCF, empirical parameters are also used to include electron correlation effects. Electron correlation effects are principally omitted in the HF method. PM7 is currently the latest development in the family of semiempirical approaches.[6, 24, 25] It offers a high accuracy of optimized geometries, thermochemistry, band gaps, and electronic spectra. PM7 favorably differs from ab initio electronic structure methods by computational cost. The performance difference comes largely from faster SCF convergence thanks to pre-parametrized integrals. PM7 routinely employs two experimentally determined constants per atom: atomic weight and heat of atomization. Electrostatic repulsion and exchange stabilization are explicitly taken into account. All applicable HF integrals are evaluated by approximate means due to computational efficiency reasons. The set of basis functions consists of one *s* orbital, three *p* orbitals, and five *d* orbitals per each atom. Basis *d* orbitals are omitted for elements without *d* electrons. The overlap integrals in the secular equation are ignored.[25]

The nuclear equations-of-motion were integrated with a 1.0 fs time-step at 300 K and with a 2.0 fs time-step at 100 K. Preliminary tests were performed to ensure acceptable total energy conservation for each integration time-step (microcanonical ensemble). The productive simulations, in turn, were performed in the constant temperature constant number of particles ensemble. The constant temperature of 300 K and 100 K (see Table 1) was maintained by the Berendsen thermostat[26] with a relaxation constant of 50 fs. The length of each nuclear trajectory (Table 1) was decided on-the-fly, depending on the convergence of results.

Table 1. The list of the simulated systems. Provided is a total number of electrons per system for a straightforward comparison of computational load with alternative electronic structure studies. Note, that PM7 uses effective-core potentials for all elements except hydrogen and helium. Equilibration time was roughly computed from the evolution of thermodynamics properties. The sampling time was set with respect to temperature, system size and convergence of results. The equilibrated molecular configurations are depicted in Figure 1

| # | # MMIM | # SCN | # $CO_2$ | # electrons | Sampling time, ps | Equilibration time, ps | Temperature, K |
|---|---|---|---|---|---|---|---|
| 1 | 4 | 4 | 2 | 360 | 180 | 20 | 300 |
| 2 | 4 | 4 | 3 | 382 | 200 | 30 | 300 |
| 3 | 4 | 4 | 4 | 404 | 200 | 30 | 300 |
| 4 | 4 | 4 | 4 | 404 | 300 | 100 | 100 |
| 5 | 0 | 1 | 1 | 51 | — | single point | 0.0 |

The latest available revision of MOPAC2012 was used to obtain wave functions and forces. The Atomistic Simulation Environment (ASE)[27] set of scripts was used as a starting point to interface electronic structure stage of computations with temperature-coupled velocity Verlet trajectory propagation. Gabedit 2.4[28] was used to visualize molecular configurations and electronic orbitals.

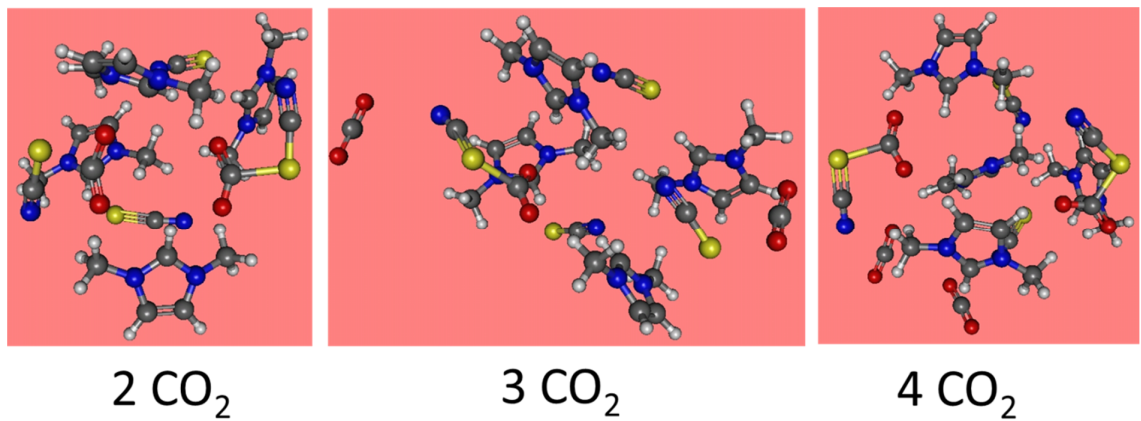

2 $CO_2$     3 $CO_2$     4 $CO_2$

Figure 1. The equilibrated molecular configurations of the simulated systems #1, #2, #3. Oxygen atoms are red, carbon atoms are grey, nitrogen atoms are blue, sulfur atoms are yellow, hydrogen atoms are white. The system compositions are provided in Table 1.

**Results and Discussion**

The equilibrated configurations of the simulated systems containing two, three, and four $CO_2$ molecules per four ion pairs of [MMIM][SCN] are depicted in Figure 1. Interestingly, certain number of $CO_2$ molecules interacts strongly with the tiocyanate anion by their carbon atoms. This binding results in a surprisingly small carbon-sulfur distance, ca. 2 Å at 300 K (Figure 2). Compare, van de Waals radius of carbon equals to 1.5 Å, whereas van der Waals radius of sulfur equals to 1.8 Å. That is, the conventional distance between these atoms must be at least 3.3 Å. It must be even larger at finite temperature due to a thermal expansion of the material. This strong spatial correlation between carbon atom of $CO_2$ and sulfur atom of the RTIL anion is dependent on the concentration of $CO_2$. The larger number of carbon dioxide is supplied, the weaker is the correlation. See heights of radial distribution functions (RDFs) in Figure 2 for quantitative details. However, some gas molecules are strongly captured at any $CO_2$ content. The remaining $CO_2$ molecules exhibit a small peak at ca. 3.4 Å. The peak height is proportional to $CO_2$ concentration. These conclusions are completely confirmed by the oxygen-sulfur RDFs, which exhibit a nearly identical trend.

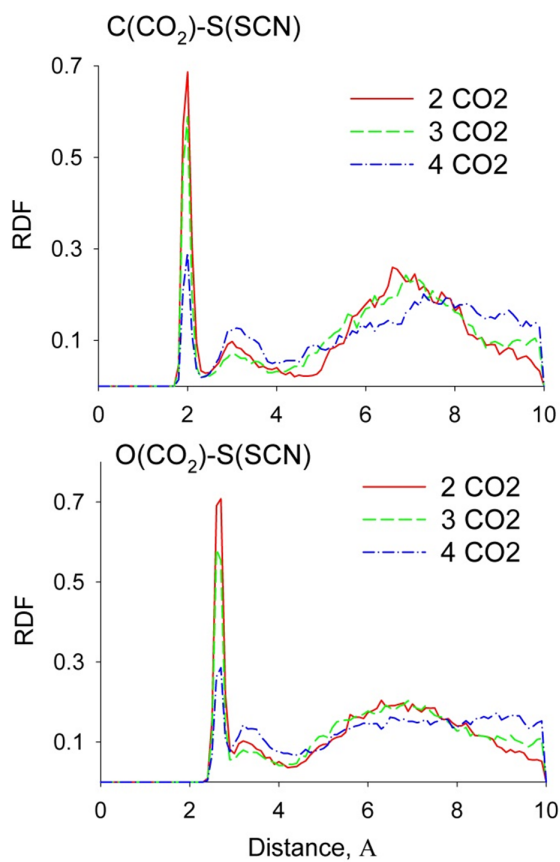

Figure 2. Radial distribution functions computed between carbon atom of $CO_2$ and sulfur atom of the anion (top); between oxygen atom of $CO_2$ and sulfur atom of the anion (bottom). The functions are normalized so that the integral of RDF equals to unity.

The same methodology was applied to characterize interactions of $CO_2$ with an imidazolium-based cation. The two most probable coordination centers for oxygen atom, imidazole ring hydrogen and methyl group carbon, are investigated in Figure 3. Interestingly, oxygen atom approaches closely to imidazole hydrogen, which is the most electron deficient site in the MMIM cation. The observed distance, 2.2 Å, qualifies for a weak hydrogen bond. It is a much weaker hydrogen bond than observed, for instance, in water, where the non-covalent oxygen-hydrogen distance falls below 2.0 Å. The height of the first peak is much smaller than in the case of $CO_2$-anion binding at the same external conditions. Compare, 0.22 to 0.69. The presence of the second and third peaks indicates some structuring, which is, most likely, a combined effect from the anion and the cation. These effects cannot be distinguished as our interaction potential is non-additive. No hydrogen bonding was detected between $CO_2$ oxygen and methyl group (Figure 3, bottom). The equilibrium distance between oxygen and carbon atoms fluctuates slightly above 3 Å. This is well consistent with their van der Waals radii.

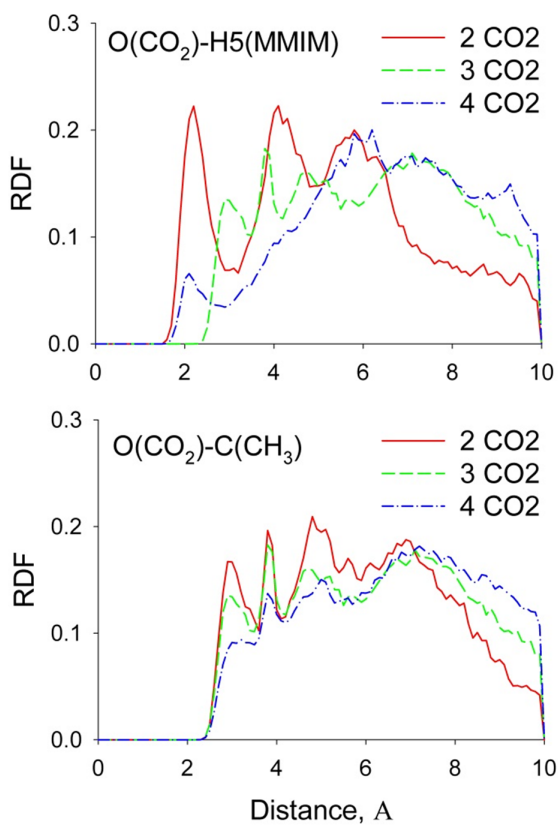

Figure 3. Radial distribution functions computed between oxygen atom of $CO_2$ and the most electron deficient hydrogen atom of the cation (top); between oxygen atom of $CO_2$ and methyl carbon atom of the cation (bottom). H5 is a common designation of the corresponding imidazole ring hydrogen atom in the common empirical force fields. The functions are normalized so that the integral of RDF equals to unity.

To recapitulate, PM7-MD calculations, supplemented by radial distributions for selected atom pairs and visual examination of immediate atomistic configurations, indicate that successful $CO_2$ capture occurs due to interaction between $SCN^-$ and $CO_2$. Although weak hydrogen bonds probably exist, thanks to the most electron deficient imidazole ring hydrogen atom, this binding is definitely weaker than that with the $SCN^-$ anion. The successful performance of the thiocyanate anion encourages trying its combinations with different cations. Such engineering may allow locating an ion pair, where entropic contribution due to gas capturing is most favorable. If such ion combination is found, the required external pressure in the industrial application may be greatly decreased. This would result in more cost efficient $CO_2$ capture/separation processes.

Thermal effects are important for the investigated phenomenon. To describe this effect quantitatively, an additional simulation (system #4 in Table 1) was performed at significantly smaller temperature, 100 K. RDFs are depicted for the same atom pairs, as in Figures 2-3, for the most straightforward comparison. The $CO_2$-cation closest contact distance shifts to 2.1 Å. This value still corresponds to a rather weak hydrogen bond. $CO_2$-anion closest contact distance remains unaltered, 2.0 Å. Generally, [MMIM][SCN]-$CO_2$ system at 100 K is, expectedly, more ordered than at 300 K. In the thermodynamic limit, this system at 100 K is in the solid state. However, the simulated system is smaller than this limit, representing rather an isolated cluster in vacuum. It is incorrect to assign this system to any macroscopic phase, since an impact of interfacial behavior is extremely significant. The determined structure, such as coordination sites and their ranking, can be smoothly extrapolated to continuous phases.

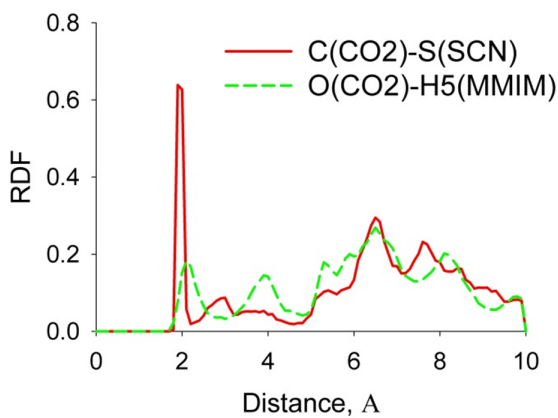

Figure 4. Radial distribution functions computed between carbon atom of $CO_2$ and sulfur atom of the anion (solid red line); between oxygen atom of $CO_2$ and the most electron deficient hydrogen atom of the cation (dashed green line). The functions are normalized so that the integral of RDF equals to unity.

Figure 5 represents selected valence electron orbitals for the [$CO_2$-SCN]⁻ complex. The most interesting feature is that the lowest unoccupied molecular orbital (LUMO, E = 0.16158 Ha) of this ion-molecular complex is well delocalized over the $CO_2$ molecule and the SCN⁻ anion. The highest occupied molecular orbital (HOMO, E = -0.04105 Ha) is localized on the SCN⁻ anion only. In turn, HOMO-1 (E = -0.04269 Ha) is slightly localized on $CO_2$, in addition to its localization on the anion. The analysis of orbital localization explains why the height of the first peak on the corresponding sulfur-carbon RDF at 100 K (Figure 4) is somewhat smaller than the height of the same peak at 300 K (Figure 2). An elevated temperature provides necessary kinetic energy for these two particles to approach one another closely and partially share certain valence orbitals. Therefore, temperature elevation from 100 to 300 K slightly favors carbon dioxide coordination by the thiocyanate anion. At the same time, $CO_2$ possesses more kinetic energy to break bonds to RTIL and create a separate phase, as argued in Ref.[19] The present results are in concordance with the insights of the above referenced work.

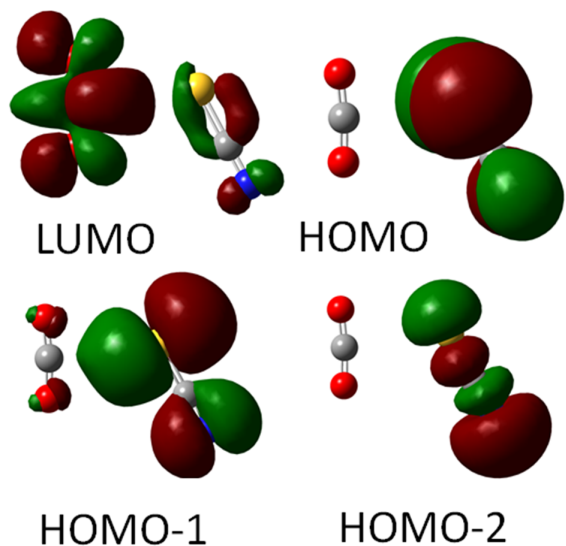

Figure 5. The localization of selected valence orbitals for the [$CO_2$-SCN]⁻ complex. Oxygen atoms are red, carbon atoms are grey, nitrogen atoms are blue, sulfur atoms are yellow. These orbitals have been calculated using highly reliable B3LYP hybrid density functional.[29, 30] The wave function has been expanded using the 6-311G(d) basis set. The coincidence of results obtained from PM7 and B3LYP is a solid proof of highly accurate results presented in this report.

**Conclusions**

This work employs the PM7-MD method to provide electronic structure level atomistic insights into the problem of $CO_2$ capture and coordination by [MMIM][SCN] room-temperature ionic liquid. The obtained result evidences that the thiocyanate anion plays a key role in gas capture, whereas the impact of cation is mediocre (in comparison). In other RTILs, the imidazolium-based cation may play a leading role due to weak hydrogen bonding ability. Decomposition of the computed wave function on the individual molecular orbitals confirms that $CO_2$-SCN binding extends beyond just electrostatic interactions in the ion-molecular system, involving partial sharing of valence orbitals.

The obtained result is in concordance with the experimental observations of performance of the thiocyanate containing ionic liquids.[19] This report provides understanding of the molecular mechanisms, which determine the reported encouraging results. PM7-MD is proven to be a powerful method to investigate many-component systems, where specific non-additive interactions take place and where propagation of molecular dynamics over hundreds of picoseconds is unavoidable. Empirical molecular dynamics simulations – although they offer a cheaper computational cost – cannot handle this problem in a reliable way without preliminary re-parametrization based on high-level ab initio computations and extensive force field evaluations. In turn, density functional based molecular dynamics is too computationally expensive to record so long trajectories. Note, that reorientation dynamics in RTILs is slow.

**Acknowledgments**

I thank Dr. James J.P. Stewart (President at Stewart Computational Chemistry, Colorado Springs, United States) for extensive discussions, wise comments and, most importantly, inspiration. I thank University of Rochester, New York, United States and personally Prof. Oleg V. Prezhdo and Dr. Eric Lobenstine for providing me a courtesy library access outside working hours. MEMPHYS is the Danish National Center of Excellence for Biomembrane Physics. The Center is supported by the Danish National Research Foundation.

I thank Nadezhda Andreeva (St. Petersburg, Russian Federation) for she knows what.

**Author Information**

E-mail address for correspondence: vvchaban@gmail.com. Tel.: +1 (413) 642-1688.